# Depressed serum IgM levels in SLE are restricted to defined subgroups

Caroline Grönwall[1], Uta Hardt[1], Johanna T. Gustafsson[1], Kerstin Elvin[2], Kerstin Jensen-Urstad[3], Marika Kvarnström[1], Giorgia Grosso[1], Johan Rönnelid[4], Leonid Padyukov[1], Iva Gunnarsson[1], Gregg J. Silverman[5], Elisabet Svenungsson[1]

1. Department of Medicine, Rheumatology Unit, Karolinska Institutet and Karolinska University Hospital, Stockholm, Sweden
2. Department of Clinical Immunology and Transfusion Medicine, Unit of Clinical Immunology, Karolinska Institutet, Karolinska University Hospital, Stockholm, Sweden.
3. Department of Clinical Physiology, Södersjukhuset, Karolinska Institutet, Stockholm, Sweden.
4. Department of Immunology, Genetics and Pathology, Uppsala University, Uppsala, Sweden.
5. Department of Medicine, Division of Rheumatology, NYU School of Medicine, New York, NY, USA

**Corresponding author:** Caroline Grönwall, *Department of Medicine, Rheumatology Unit, Karolinska Institutet and Karolinska University Hospital, Center for Molecular Medicine L8:04, 17176 Stockholm, Sweden*. caroline.gronwall@ki.se

## Abstract
Natural IgM autoantibodies have been proposed to convey protection from autoimmune pathogenesis. Herein, we investigated the IgM responses in 396 systemic lupus erythematosus (SLE) patients, divided into subgroups based on distinct autoantibody profiles. Depressed IgM levels were more common in SLE than in matched population controls. Strikingly, an autoreactivity profile defined by IgG anti-Ro/La was associated with reduced levels of specific natural IgM anti-phosphorylcholine (PC) antigens and anti-malondialdehyde (MDA) modified-protein, as well total IgM, while no differences were detected in SLE patients with an autoreactivity profile defined by anti-cardiolipin/$\beta_2$glycoprotein-I. We also observed an association of reduced IgM levels with the HLA-DRB1*03 allelic variant amongst SLE patients and controls. Associations of low IgM anti-PC with cardiovascular disease were primarily found in patients without antiphospholipid antibodies. These studies further highlight the clinical relevance of depressed IgM. Our results suggest that low IgM levels in SLE patients reflect immunological and genetic differences between SLE subgroups.

## Keywords:
SLE, natural IgM, Sjögren's syndrome, antiphospholipid syndrome, IgM anti-PC, IgM anti-MDA, anti-CWPS, SS, APS, autoantibodies

## Highlights:
* Low IgM levels are more common in SLE patients than in controls
* Patients with anti-Ro/La autoimmunity often have strikingly decreased IgM levels
* Low IgM was seen for specific IgM anti-PC, anti-MDA, anti-CWPS, as well as total IgM
* Low natural IgM may be associated to HLA-DRB1*03 in both SLE and controls

## 1. Introduction
Systemic lupus erythematosus (SLE) is an autoimmune disease associated with a great diversity of clinical features, complex pathogenesis [1, 2], and a central hallmark of distinct types of IgG autoantibodies. Almost all SLE patients are at the time of diagnosis positive for antinuclear autoantibodies (ANA), an umbrella immunofluorescence test that detects nuclear specificities, which include disease-specific IgG anti-dsDNA antibodies, present in 40-60% of SLE patients. Such IgG lupus autoantibodies have been hypothesized to be directly pathogenic in SLE through pathways mediated by Fc-receptor interaction, TLR activation, and induction of type I interferons [3, 4]. Yet, it remains controversial whether all autoantibodies are directly pathogenic or represent bystander effects that are peripheral to the disease process. Indeed, lupus is associated with a number of IgG-autoantibody specificities including anti-Sm, anti-RNP, anti-C1q, anti-nucleosome, anti-Ro/SSA, anti-La/SSB and antiphospholipid autoantibodies (i.e. anti-CL/$\beta_2$GPI) [1, 2].

Some SLE patients are affected by secondary clinical syndromes, i.e. secondary Sjögren's syndrome (sSS) and secondary anti-phospholipid syndrome (sAPS), which display distinct clinical features. APS patients suffer from arterial and venous thrombotic events and/or pregnancy morbidities [5-7], and SS patients have involvement of lachrymal and salivary glands [8, 9]. For secondary APS, anti-CL/$\beta_2$GPI autoantibodies are included in the diagnosis [6] while for secondary SS the presence of primary SS-associated autoantibodies (anti-Ro/La) are not required [9].

The appearance of diverse autoreactivities in SLE patients has been suggested to result in part from defects in the homeostatic clearance of apoptotic cells leading to exposure to apoptotic cell antigens in an inflammatory milieu [10]. Intriguingly, natural IgM often recognize epitopes expressed on apoptotic cells. These IgM are present already at birth, and are postulated to convey beneficial housekeeping properties, which may also enhance apoptotic cell clearance [11, 12].

Natural IgM are primarily constitutively produced by specialized innate like B-cell subsets, without the need for additional immune stimulation. Indeed, in mice up to 80% of the circulating IgM molecules are spontaneously produced [13, 14]. The repertoire of natural IgM, in both mice and humans, is biased towards the recognition of oxidation-associated antigen-specificities, which include epitopes on malondialdehyde (MDA) oxidation-modified proteins, and phosphorylcholine (PC) in oxidized lipids [15]. PC epitopes are present on apoptotic cells, in oxidized low-density lipoprotein, and also on the *pneumococcal* cell wall polysaccharide. Natural IgM anti-PC may thus have dual roles in both enhancing phagocytic clearance of dead cells, and as a first line of defense against invasive infections, including to *S. pneumoniae*.

Importantly, IgM anti-PC antibodies have been shown to be directly anti-inflammatory in *in vitro* assays and *in vivo* in murine models [16-20]. A large number of studies have demonstrated that low levels of IgM anti-PC are associated with higher risk of cardiovascular disease both in SLE patients and in the general population [21-28]. It can be speculated that there is a dynamic balance between these protective IgM and disease-associated pathogenic autoreactive IgG in SLE. Hence reduced levels of circulating IgM may by itself significantly affect SLE pathogenesis. Yet, SLE is an immensely heterogeneous disease and there are often great differences between patients. We therefore hypothesized that IgM levels may differ between subsets of patients with SLE. In the current study, we document that there are large variations in the levels of total IgM and in specificities of certain natural IgM antibodies in SLE subgroups defined by lupus IgG autoantibody profiles.

## 2. Materials and methods

### 2.1 Subjects, clinical data, and subgroups
This cross-sectional study includes 437 consecutive patients from the Karolinska SLE cohort enrolled between 2004-2011 and fulfilling at least four of the 1982 revised classification criteria for SLE according to the American College of Rheumatology [29]. SLE patients were assigned to subgroups based on autoantibody profiles at the time of blood sample collection into antiphospholipid (APS)-profile or Sjögren (SS)-profile (Table 1). The patients assigned to the APS-profile group had either two or more IgG/IgA/IgM anti-CL/$\beta_2$GPI positive tests, and/or a lupus anticoagulant (LA) positive test. The SS-profile group had to have two or more positive IgG anti-Ro52/Ro60/La tests. Notably, clinical diagnosis and manifestations were not considered in the subgrouping. We focused exclusively on current immunological autoreactivity status. The following patients were excluded: patients that would fulfill both profile criteria (n=19), patients that did not have sufficient serological tests for subgrouping (n=42), and patients that had received Rituximab treatment at any time point prior to blood sampling (n=40). The remaining SLE patients were subgrouped as having another autoantibody profile ("other Abs"). Disease activity was measured by the SLE disease activity index 2000 (SLEDAI-2K)[30] and SLE associated organ damage was assessed by the Systemic Lupus International Collaborating Clinics (SLICC)/ACR Damage Index [31]. For comparison of distribution between subgroups clinical assessment of secondary APS was determined by fulfillment of clinical criteria by Miyakis et al. [6] and secondary SS by the American European Consensus criteria [9]. Atherosclerosis was evaluated with carotid ultrasound intima-media thickness (IMT) using a duplex scanner with a linear array transducer. Both right and left carotids were measured and presence of plaque was defined as ≥100% increase over background in any arterial segment [32].

Serum samples were obtained from 322 population controls and 287 of these samples were individually age and sex matched to 287 SLE patients for comparison of IgM levels. A majority of individuals were also matched for residential area. Control subjects were identified in the Swedish national population registry without influence of clinical history and invited to participate. The only exclusion criterion was a diagnosis of SLE. All study participants were examined by a rheumatologist at the Karolinska University Hospital and medical records were reviewed. Validation assays also included 597 cryopreserved population control serum samples with known HLA-DRB1 alleles collected within the Epidemiological Investigation of Rheumatoid Arthritis (EIRA) study [33]. The local ethic committee at Karolinska University

Hospital approved the study and all the experiments were performed according to good clinical practice and good laboratory practice. All study subjects gave written informed consent to participate.

### 2.2 Antibody measurements
Total IgM and total IgG levels were measured at the Karolinska University Hospital clinical laboratory by standard nephelometry. Antibodies to specific nuclear antigens (dsDNA, SSA/Ro52, SSA/Ro60, SSB/La, Sm, RNP) and antiphospholipid IgG, IgA and IgM (cardiolipin and $\beta_2$-glycoprotein-I) were analyzed by multiplexed bead technology (Luminex) using BioPlex 2200 system (Bio-Rad, Hercules) according to the specifications of the manufacturer. Analysis of IgM rheumatoid factor (RF) was performed on a Phadia ImmunoCap 250 instrument according to the manufacturer's instructions. LA was determined using a modified Dilute Russell Viper Venom method (Biopool) using Bioclot lupus anticoagulant.

Serum levels of natural antibodies to malondialdehyde (MDA)-modified protein adducts or phosphorylcholine (PC) were measured by sandwich ELISA using previously described methods [25, 26]. High-binding half-area ELISA plates (Corning) were coated with PC6-bovine serum albumin (PC-BSA, Biosearch Technologies), pneumococcal cell wall polysaccharide CWPS (CWPS, Statens Serum Institut), or MDA-modified BSA, (MDA-BSA, Academy Biomedical) at 3 μg/ml in PBS and blocked with 3% BSA in PBS. Samples were analyzed at 1:200 and 1:1000 dilution. Binding was detected with HRP conjugated mouse anti-hIgM (μ-chain specific, Southern Biotech) followed by TMB substrate development. Absorbance was normalized to a standard curve of a positive reference sample which was included in all assays and quantitative reactivities were presented as Relative Units/ml (RU/ml).

### 2.3 HLA-DRB1 profiling
Low resolution HLA-DRB1 genotyping was performed by sequence-specific primer PCR as previously described [34]. The following HLA-DRB1 allelic groups were detected by this method: DRB1*01, DRB1*03, DRB1*04, DRB1*07, DRB1*08, DRB1*09, DRB1*10, DRB1*11, DRB1*12, DRB1*13, DRB1*14, DRB1*15 and DRB1*16, and were used for statistical evaluation with dominant model.

### 2.4 Statistical analysis
The distribution of antibodies, demographic parameters, and biomarkers were compared between different groups with two-sided Mann-Whitney or Kruskal-Wallis test, as appropriate. Frequency distributions were compared with Fisher's exact test. Non-parametric Spearman correlation was used to analyze associations between variables. Logistic regression models were used for multivariate analysis. A p-value of less than 0.05 was considered significant. Analysis was performed using Prism 6 (Graphpad Software Inc), JMP 13 (SAS Institute Inc) and Review Manager 5.3 (Copenhagen: The Nordic Cochrane Centre, The Cochrane Collaboration, 2014).

## 3. Results

### 3.1 The frequency of depressed IgM is higher in SLE patients than in matched controls
We initiated our studies by investigating whether natural IgM levels were altered in SLE patients compared to unaffected population controls. Two different classes of natural IgM antibody specificities were studied, anti-PC and anti-MDA modified protein epitopes. While these are distinct types of oxidation-associated epitopes, specific anti-PC antibodies are cross-reactive with PC-epitopes present both in lipids and bacterial polysaccharide. In our studies, we therefore measured PC-reactivity with two different assays; using the small PC headgroup conjugated to a carrier protein (PC-BSA), or purified cell wall polysaccharide (CWPS).

Our results show that average serum total IgM levels were higher in SLE patients compared to age- and gender- matched population controls, although the differences were very small, and the median value was lower in the SLE patients (SLE vs controls: mean±SD: 1.276±1.1 vs 1.265±0.71 mg/ml; median: 0.96 vs 1.00 mg/ml, Table 2, Figure 1). Similarly, the average level of natural IgM binding to oxidation-associated MDA-modifications was higher in SLE patients than in controls, while the median value was lower. In contrast, the levels of natural IgM anti-PC were lower overall in the SLE cohort compared to age- and gender- matched controls. Indeed, by both independent assays, which measured levels of IgM reactive with PC-conjugate or with CWPS, there were similar findings (Table 2, p=0.05 and p<0.0001). Notably, IgM anti-PC-BSA and IgM anti-CWPS showed a strong direct correlation (p<0.0001, R=0.85, Supplemental Figure 1). Importantly, both from analysis for total IgM, as well as for anti-PC-BSA, anti-CWPS, and anti-MDA antibodies, there were more SLE patients with depressed IgM levels compared to the controls (Table 2). These patterns were not solely a reflection of total IgM levels, and highlight the heterogeneity of the SLE population.

In our studies, we defined depressed specific IgM and total IgM in the SLE patients by a cutoff at the 25th percentile of the population controls and high IgG by a cutoff at the 75th percentile of the population controls. For total IgM, all tested individuals had detectable levels of IgM, and most of the SLE patients with low IgM were still within the normal range (>0.46mg/ml), based on the reported 2.5th percentile a study of the general population (Mean: 1.47± 0.84 mg/ml) [35]. In fact, only a small proportion, 13.6% of the SLE compared to 3% of the unaffected controls (of 322 subjects tested), had subnormal IgM levels. For IgG, mean levels of total IgG levels were significantly elevated in SLE patients (p<0.0001), and 60% had high IgG levels, over the 75% population control cutoff.

We found significant direct correlations between each of the specific natural IgM antibody levels and total IgM (p<0.0001, Supplemental Figure 1), which confirmed findings in a previous report [26]. Notably, there were significant but modest inverse correlations of IgM anti-PC-BSA, anti-CWPS and anti-MDA with age (p=0.0006, R=-0.16; p<0.0001, R=-0.25; p=0.0007, R=-0.17, respectively) (Supplemental Figure 2). IgM anti-MDA demonstrated a weak positive association with disease activity by SLEDAI [30], while IgM anti-PC-BSA and IgM anti-CWPS correlated inversely with organ damage by SLICC [31] (Supplemental Table 1). The IgM levels did not correlate with current glucocorticoid treatment dose (Supplemental Table 2). As SLE is a female predominant disease, it may be important to note that total IgM levels and IgM anti-MDA were both significantly lower in men than women, among both SLE patients and unaffected population controls (controls: p=0.03 and p=0.03, respectively; SLE: p=0.04 and p=0.03, respectively). IgM anti-PC levels were also lower in the male SLE patients compared to female patients (p=0.03, Supplemental Table 3, Supplemental Figure 3). IgM anti-PC levels were in addition analyzed in a second independent population control cohort with a different demographic distribution (i.e., 27% males). Here, we confirmed that IgM anti-PC levels are significantly lower in male than in female adults (average 23% lower levels, p=0.0002) (Supplemental Figure 4). Yet, stratifying the patients based on sex did not markedly change the correlation between IgM levels in SLE patients compared to controls (Supplemental Table 4).

### 3.2 IgM levels are primarily decreased in SLE patients with an IgG anti-Ro/La autoreactivity profile

We next investigated whether natural IgM levels varied between different SLE subsets based on autoantibody profiles. Patients were characterized as having an APS-profile (with anti-CL/$\beta_2$GPI) or a SS-profile (with IgG anti-Ro/La) as defined in Table 1. The remainder were assigned to a third group, SLE with other autoantibodies. Patients in this group often displayed anti-Sm/RNP autoantibodies. Indeed, these three subgroups were generally mutually exclusive, and only a few patients had either overlapping profiles or the data were insufficient for assignment. These three SLE patient subsets did not differ regarding mean age or disease activity (Table 3). However, the APS-profile subset had more anti-dsDNA positive patients than the SS-profile group (50% vs 28%, p=0.005).

Among the most striking characteristics for the patient subset with the SS autoantibody profile were the significantly lower total serum IgM levels, as well as low IgM anti-PC-BSA, IgM anti-CWPS and IgM anti-MDA levels (Table 3, Figure 1). In contrast, patients with the APS autoantibody profile displayed higher mean levels of total IgM and IgM anti-MDA compared to controls, although these differences were not statistically significant. In the third SLE group (SLE with other autoantibodies) we also observed significant depression of some of the IgM measurements, although these associations were not so strong as for the SS subgroup. Total IgM seems an important component driving these correlations. However, in SS-profile SLE patients, IgM anti-PC-BSA and IgM anti-CWPS levels were still significantly lower after normalizing for total IgM levels (i.e., IgM anti-PC-BSA/total IgM; IgM anti-CWPS/total IgM), whilst IgM anti-MDA/total IgM was not significantly different (Supplemental Table 5, Supplemental Figure 5). This may suggest that certain IgM specificities are more affected than others by the patients autoreactivity profile, and it also emphasizes that there are inherent differences between IgM anti-PC and IgM anti-MDA expression patterns. When dichotomizing the cohort base on only one autoantibody test, lower associated levels of IgM were observed based on anti-Ro60 positivity than anti-Ro52 positivity, although not as remarkable as when using the autoantibody profiling strategy (data not shown). Hence, IgG anti-Ro60 positivity may be the stronger driver for low IgM in the SS-profile. No significant differences between IgM levels were observed after dichotomization of SLE patients according to their IgG anti-dsDNA status (Supplemental Figure 6) and the depressed IgM levels in the SS group were observed independently on anti-dsDNA status (data not shown). We also confirmed previous findings that IgM anti-PC levels are not affected by lupus nephritis [25], while the total IgM levels were lower in patients with a history of renal disease (Supplemental Table 6). Moreover, not all types of IgM specificities were equally depressed in the SS subgroup. In fact, levels of disease-associated rheumatoid factor IgM (RF IgM) were significantly elevated in the SS-profile group, which may reflect the effects of a specific immune stimulation in this subset (Supplemental Figure 7, Supplemental Table 7). Similarly, while there was generally a direct correlation between in RF IgM and other IgM levels when all SLE patients were analyzed (Supplemental Figure 7), the few patients with high RF titers all had relatively low specific natural IgM levels and total IgM. These findings also confirmed that our assays were without technical bias caused by the presence of rheumatoid factors. Taken together, depressed IgM levels were primarily associated with the subgroup of SLE patients with SS autoantibody profile.

To understand whether there were relationships between patients based on conserved patterns of (auto)antibody expression, we performed unsupervised hierarchical cluster analysis (using Cluster 3.0), (Figure 2). Our findings demonstrated that depressed IgM levels (i.e., both total IgM and specific natural IgM), enabled visualization of more general effects on heterogeneity of lupus antibody expression patterns. These analyses independently confirmed that low IgM levels were commonly found in the subgroup of SLE patients with anti-Ro/La IgG autoantibodies. In the cluster analysis, specific variations in HLA-DRB1 allelic expression also contributed to the discontinuous variations represented as different IgG autoantibody defined patient subgroups. Indeed, the three patient subsets displayed differences in the distribution of the HLA-DRB1 alleles, as the APS-profile group displayed a higher frequency of HLA-DRB1*04 (p=0.002), while the SS-profile was enriched for HLA-DRB1*03 (p<0.0001). The remaining patients, that included many subjects with anti-Sm/RNP IgG autoantibodies, more commonly had HLA-DRB1*15 alleles (p=0.0005) as well as HLA-DRB1*03 (p=0.001) alleles compared to the control group (Table 3). To a large extent, these findings confirm previously reported HLA-DRB1-autoantibody associations in SLE [34, 36, 37].

### 3.3 Low IgM levels may be associated with HLA-DRB1*03 alleles

As we observed strong HLA associations with IgG autoantibody profiles, and considering that the anti-Ro/La subgroup had both lower IgM and was significantly enriched for HLA-DRB1*03, it was not unexpected that there was also a significant association between low IgM and HLA DRB1*03. Indeed, SLE patients carrying a HLA-DRB1*03 allele had lower total IgM (p=0.05), IgM anti-PC-BSA (p=0.02), and IgM anti-MDA (p=0.05), as well as a trend towards lower IgM anti-CWPS (p=0.07) (Table 4, Figure 3). More surprisingly, we also found lower IgM levels in the HLA-DRB1*03 positive population controls, which were statistically significant for total IgM levels (p=0.03) and IgM anti-PC-BSA levels (p=0.04) (Table 4, Figure 3). These same associations were strengthened when both SLE and controls were combined in a meta-analysis, which revealed that HLA-DRB1*03 positive individuals had lower total IgM (p=0.007), IgM anti-PC-BSA (p=0.002), IgM anti-CWPS (p=0.007), and IgM anti-MDA levels (p=0.02) (Table 4, Figure 3). Similarly, in the groups with HLA-DRB1*03 positivity, the frequency of individuals with low IgM levels was also significantly higher (Table 4, Supplemental Figure 8). In addition, the control group showed trends towards allele copy dose-dependency as individuals homozygous for HLA-DRB1*03 displayed lower IgM levels seen in heterozygous, or null individual, with a significant difference for total IgM and IgM anti-CWPS levels in the HLA-DRB1*03 (-/-) group (p=0.03, p=0.04, respectively) (Supplemental Figure 9, Supplemental Table 8). Yet, when analyzing IgM anti-PC levels in a second independent population control cohort we did not find statistically significant differences in IgM levels between DRB1*03 negative, heterozygous, and homozygous individuals, although the numeric trends remained consistent (Supplemental Figure 10).

By contrast, in these analyses there were no correlations between HLA-DRB1*03 and total IgG levels, in SLE or control groups (Table 4). We also investigated other HLA-DRB1 variants, and in the meta-analysis that combined results from SLE patients and controls we documented significantly higher IgM anti-PC levels in individuals with HLA-DRB1*04 positivity (p=0.009, Supplemental Table 9). In addition, the SLE patients with HLA-DRB1*15 had somewhat lower total IgG levels (p=0.05) but there was no association with levels of IgM (Supplemental Table 10).

In multivariate logistic regression analysis, IgM levels remained significant independent variables when adjusting for DBR1*03, DRB1*04, DRB1*15, sex, age, and RF IgM positivity in models for determining if the SLE patient had the APS-profile, the SS-profile, or neither (IgM anti-PC-BSA p=0.03, IgM anti-CWPS p=0.03, IgM anti-MDA p=0.0001, total IgM p=0.0005) (Supplemental Table 11). These data may indicate that differences in IgM levels are influenced by other immunobiological co-factors.

### 3.4 Low IgM associated cardiovascular risk is different in SLE subgroups

SLE patients are reported to have an increased risk of premature cardiovascular disease, compared to matched unaffected controls [38, 39], and this risk may be greater in the subgroup of SLE patients with anti-phospholipid antibodies, while the risk is lower in SLE with anti-Ro/La positivity [40]. In our analysis, we confirmed a higher odds ratio (OR) for a history of a vascular event, including both arterial and venous events, in the APS-profile group (OR=2.7 CI: [1.6-4.4] p=0.0003), and a significantly lower OR for vascular events in the SS-profile group (OR=0.46 CI: [0.24-0.87] p=0.01), compared with the overall frequency in SLE patients (Table 5).

When analyzing the complete SLE cohort, we observed an overall association for the presence of atherosclerotic carotid plaque with low IgM anti-PC (Table 5). SLE patients with low IgM anti-PC-BSA had an OR of 2.2 CI:[1.2-4.0] of having detectable carotid plaques by ultrasound (p=0.02) and patients with low IgM anti-CWPS similarly had an OR of 3.5 CI:[1.9-6.5] in association to plaque (p<0.0001). Importantly, even after adjusting for age and sex, the frequency of low IgM anti-CWPS in association to plaque remained significant (p=0.005). Notably, the association between low IgM anti-PC and cardiovascular disease was relatively moderate in this study compared to other cohorts [25, 26].

Vascular events were significantly more prevalent in the APS-profile subgroup (48% compared to 23% in the total SLE cohort, p=0.0003), and were not influenced by depressed IgM anti-PC levels. In contrast, there was a significantly higher frequency of vascular events with low IgM in the "other autoAbs" group (low IgM anti-PC-BSA, OR 2.5, p=0.04). Similarly, there was no correlation between low IgM anti-PC and plaques in the APS-profile, while the frequency of atherosclerotic plaques in with low IgM anti-CWPS reactivity were significantly more prevalent in both the SS-profile group and the "other Abs group" (low IgM anti-CWPS p=0.0003 and

p=0.002) (Table 5). Results from the PC-BSA based assay, appeared to detect a higher frequency of patients with carotid plaques with low IgM anti-PC in the SS-profile SLE subgroup (28% vs 5%, OR=7.2, CI:[1.5-36], p=0.01) (Table 5). Consequently, while plaques are less common in the SS group, low IgM anti-PC levels may have a stronger impact on the atherosclerosis risk in this group. Furthermore, logistic regression analysis with both the autoantibody profile, sex, age, and low IgM levels, demonstrated that either low IgM anti-CWPS (p=0.003) or low IgM anti-PC-BSA (p=0.02) were significantly associated with presence of plaque, while total IgM and IgM anti-MDA were not (Supplemental Table 12).

Cumulatively, our investigations have demonstrated that the immunologic heterogeneity among SLE patients is in part intertwined with low circulating levels of IgM, which may have differential effects on disease manifestations in individual SLE patients.

## 4. Discussion

Our study evaluated the expression levels in three lupus-antibody defined SLE subgroups of IgM that recognize oxidation-associated epitopes, PC and MDA, which are common components of the natural antibody repertoire that is postulated to have protective homeostatic properties. We demonstrated that the frequency of individuals with depressed levels of total IgM, and of either the tested oxidization-associated IgM specificities, is more prevalent in SLE patients than in unaffected matched controls. Strikingly, we documented large differences in IgM levels between the three SLE subgroups, with dramatically lower IgM levels observed in patients with IgG anti-Ro/La autoreactivity, compared to patients with antiphospholipid autoantibodies. However, the mechanisms responsible for these altered patterns of IgM expression in autoimmunity are essentially unknown. Furthermore, the higher probability of atherosclerotic cardiovascular disease associated with low IgM is primarily associated with the patients without antiphospholid antibodies. Our investigation therefore highlights the heterogeneity of SLE patients based on their autoantibody profiles, especially between anti-Ro/La versus anti-CL/$\beta_2$GPI profiles.

We have previously reported that anti-PC and anti-MDA recognize distinct epitopes and these antibodies display non-overlapping binding specificities [25, 26, 41]. Especially IgG anti-PC and anti-MDA have very different expression pattern in autoimmune patients and in rheumatoid arthritis IgG anti-MDA levels are associated with higher disease activity and potentially pathogenic properties [42]. IgM anti-MDA is a natural antibody specificity that is prevalent at birth but may also arise in part in response to inflammatory stimuli that are increased during autoimmune pathogenesis [25, 26, 41]. On the other hand, IgM anti-PC, are more prevalent in the circulation of healthy adults. However, anti-PC antibodies can also recognize determinants on the cell wall polysaccharide of all *S. pneumonia* strains and therefore their levels may be further induced in response to bacterial exposure. Murine studies have suggested that there are differences in PC-recognition dependent on the antigen, consequently leading to different classes of anti-PC antibodies [43]. To ensure completeness in our serologic surveys, we therefore decided to measure two different forms of phosphorylcholine, the PC headgroup conjugated to a carrier (PC-BSA) and the PC epitope in pneumococcal cell wall polysaccharide (CWPS). Patterns of reactivity to these two antigens were very similar, although not identical, and we conclude that there is, as expected, a large overlap between oxidized lipid- and polysaccharide-reactivity in humans.

We speculate that certain IgM binding specificities may be inducible by increased oxidation, apoptotic cells, and overall burden of neo-antigens, while others may be more dependent on genetic predisposition. Moreover, the balance between expression and consumption of natural IgM is not understood. In this report, we use the terminology "depressed IgM" and postulate that IgM levels are affected by active processes, yet longitudinal studies are needed to reveal if low IgM precedes lupus or is a consequence of the disease process. IgM generally have short circulating half-lives of at most a few days, and unlike IgG there is no mechanism for IgM re-cycling. As a consequence, antigen recognition and binding likely leads to consumption and clearance. In studies that integrated the sorting of PC-specific peripheral B cells, serum levels of IgM anti-PC antibodies were shown to be proportional to the frequency of PC reactive-circulating B cells [44]. Moreover, a cohort of SLE patients was shown to have lower serum total IgM compared to healthy controls, with higher proportions of polyreactive IgM [45]. We therefore postulate that in some patient groups, there are increased levels of certain specificities of autoreactive antibodies in the IgM repertoire.

The observed differences in the levels of IgM to oxidation-associated determinants were relatively modest in the comparisons of the complete SLE cohort with the matched population controls. Importantly, our studies also showed that IgM levels are generally higher in women than men, and lower in older individuals, which further highlights the importance of well-matched control cohorts. The inverse correlation of natural IgM levels with age may reflect changes in the natural antibody repertoire, and there are distinct phases of immune-development occurring in the perinatal period, as well as later age-associated changes, of innate-like B cells [46]. Although also previously reported, the factors responsible for higher levels of certain natural IgM binding specificities in women remain obscure [27, 47, 48].

We hypothesize that the dramatic differences in IgM levels in the SLE subgroups reflects variations in the underlying immunopathology in these distinctly different patients. Of the several possible explanations for low natural IgM, it may reflect consumption of IgM that follows immune complex formation with the increased burden of apoptotic cell breakdown products, and/or a disease-associated shift in the B-cell repertoire away from IgM secretion and towards IgG production. Primary Sjögren's patients are known to have very high circulating IgG anti-Ro/La autoantibody levels and elevated total IgG [49]. Indeed, in our study the patients with SS-profile had higher total IgG levels, although there was no direct correlation between IgM and IgG levels. Alternatively, a generalized pro-inflammatory milieu may directly affect the B cells themselves, as activated leukocytes are a source of B-cell survival factors that interfere with immune tolerance deletion mechanisms as well as enhanced differentiation and autoantibody expression. Stimulation of innate immune receptors, such as TLR7/9, can besides promoting IgG class-switch, also induce high level IgM production [50]. Hence, TLR ligands and/or cytokine expression likely differentially affecting the different SLE subgroups. Furthermore, a recent report demonstrates depressed IgM levels also in female first-degree relatives to SLE patients, further suggesting a role of genetic, epigenetic, or shared environmental factors [51]. It is also important to acknowledge that the ethnicity distribution in study cohorts may influence the underlying genetics and autoantibody profiles, and it is possible that the European SLE population presented here may differ with regard to antibody profile and subgroups as compared to other SLE cohorts with patients of different ethnicities. Consequently, IgM levels may be dependent on genetic influences, predisposing for production of certain types of protective (or pathogenic) autoantibodies.

In general, MHC allelic variation is the strongest genetic contribution to SLE susceptibility, and especially HLA-DRB1*15 and HLA-DRB1*03 that have documented an overall association with SLE diagnosis [52, 53]. In particular, HLA-DRB1*03 appears to predispose to development of secondary Sjögren's syndrome and anti-Ro/La positivity [36, 37]. Similarly, HLA-DRB1*04 and HLA-DRB1*13 are reported to correlate with increased risk for antiphospholipid IgG positivity and associated vascular events in SLE [34]. In the current study, we found that a high proportion of the subgroup with APS-profile bore HLA-DRB*04, while the SS-profile group displayed high representation of HLA-DRB1*03. We also found that IgM levels were generally lower in the SS group, and in an individual DRB1*03 inheritance also correlated with decreased serum IgM levels. Unexpectedly, this correlation was also documented in the population controls that were negative for anti-Ro/La IgG. The results were confirmed in our meta-analysis, which combined the SLE cohort and the population controls. We hypothesize that HLA-DRB1 alleles may affect serum total IgM levels, which we found are preferentially associated with different IgG autoantibody-defined SLE subsets, although other factors are also likely contributory. While the correlation of specific HLA alleles with IgG autoantibodies may be explained by preferential autoantigen presentation and T cell activation, a correlation between natural IgM with HLA has not been previously investigated.

Our data propose that IgM levels may be partly genetically determined. Natural antibodies are generally defined as spontaneously secreted T-cell independent responses, which primarily arise from innate-like B-1 cells [13]. Yet, it has recently been reported that anti-PC and anti-MDA, considered prototypic types of natural IgM antibodies, can at least to some extent, be T-cell dependent [54]. Indeed, innate-like B cells can also serve as antigen-presenting cells for T cells [55]. Importantly, IgM anti-PC, which can protect from disseminated pneumococcal infection, are impaired in mice transgenic for HLA-DRB1*03 [56]. However, the biochemical and immunological biases for these impaired post-immune responses remain unexplained.

IgG anti-Ro autoantibodies are associated with neonatal lupus and congenital heart block in a subset of pregnancies in women with anti-Ro autoimmunity [57, 58]. Our previous studies of newborn babies in a neonatal lupus cohort showed that IgG anti-Ro autoreactivity in the mother can affect the natural IgM levels in the baby [41]. In fact, if the mother carried IgG anti-Ro autoantibodies, we observed that the baby had higher levels of total IgM but decreased levels of natural IgM anti-MDA. Notably, these findings were neither dependent of symptoms of autoimmune disease in the mother nor the child [41]. This association may suggest a change in the B cell repertoire leading to reduced production of some IgM and higher expression of others. This may be due to anti-Ro IgG crossing the placenta and forming stimulatory immune complexes in the fetal circulation, or, alternatively, fetal B cells may be stimulated by other pro-inflammatory molecules from the mother. Yet, it may also reflect shared inherited factors, that later in life contribute to an increased risk of developing IgG anti-Ro/La autoreactivity. Consequently, the low IgM levels in the SS-profile group observed in the current studies could have first arisen during the perinatal period. Furthermore, even though murine studies have convincingly demonstrated that autoimmune disease is significantly augmented by lack of circulating polymeric IgM [59], other studies have shown that under certain circumstances natural IgM can also induce an interferon signature [59]. Hence, we do not know how differences in the natural IgM may affect the development of autoantibodies in humans.

Many reports have shown that low natural IgM to PC, and to certain MDA-modified epitopes, are associated with higher risk of atherosclerosis and cardiovascular events in SLE [21, 25, 26, 28, 60]. Here we report the lowest IgM anti-PC levels in the anti-Ro/La patient subgroup and higher in the APS-profile group despite the fact that the latter bear a higher risk of vascular events [40]. The increased occurrence of cardiovascular events in SLE is believed to arise from the contributions of different underlying pathologic influences, including antiphospholipid antibody associated thrombosis, impaired renal function, myocardial perfusion abnormalities, endothelial changes and accelerated atherosclerosis [61]. Low protective IgM anti-PC levels are more likely to contribute to processes responsible for vascular inflammation and atherosclerosis, which is consistent with our results. Moreover, it may be important to consider these differences when using low anti-PC as marker of cardiovascular risk. The results may be significantly influenced by the patient subgroup, and possibly genetic, immunological, and sex differences between patients with different autoantibody predominance.

In summary, our current findings demonstrate that depressed IgM levels in SLE, both total and natural IgM, are common in the SS-profile subgroup of the SLE patients, which are also predominantly HLA-DRB1*03 positive individuals. In contrast, SLE patients with an APS-profile had similar IgM levels as the controls. Whereas SLE is an extremely heterogeneous disease with patients that vary greatly by immunological profiles and clinical manifestations, and our new perspectives on depressed natural IgM levels provide an entirely new set of insights into the determinators of patient immunological and clinical diversity. This study is a further step towards clarifying the relationship between natural IgM and IgG autoreactivity. Yet, much remains to be learned about the protective role of IgM in relation to the early development of autoimmune disease. Similarly, we still need to increase our knowledge about how genetic and environmental factors may shape the B cell repertoire and contribute to predisposition to certain autoimmune manifestations.

## Acknowledgements

We thank Lena Israelsson, Eva Jemseby, and Julia Boström for managing the cohorts biobanking and handling of blood samples. This work was supported by the Swedish Research Council, Åke Wiberg's foundation, Professor Nanna Svartz foundation, Magnus Bergvall's foundation, Swedish Heart-Lung foundation, Stockholm County Council (ALF), King Gustaf Vs 80th Birthday Fund, Swedish Rheumatism Association, Swedish Society of Medicine, Karolinska Institutet's Foundations, the Judith and Stewart Colton

Foundation, and the foundation in memory of Clas Groschinsky.

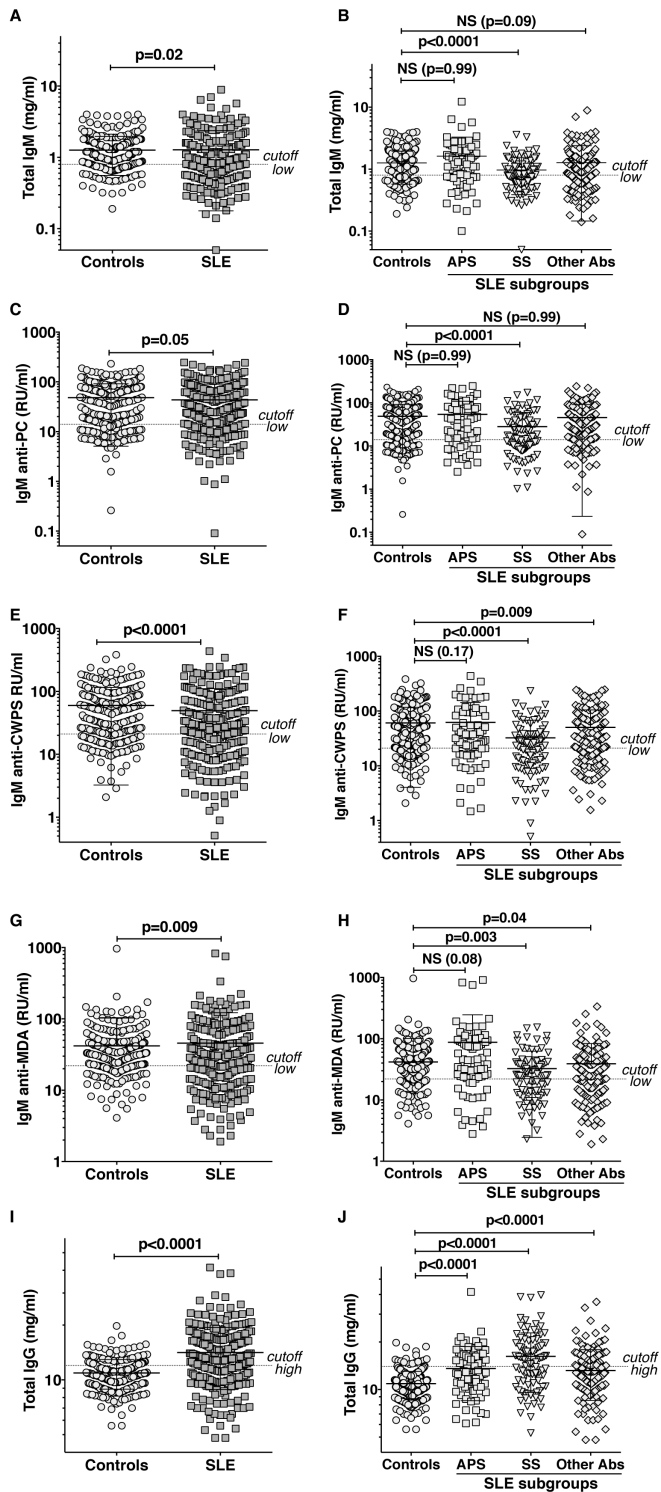

**Figure 1. IgM levels in different SLE subgroups**

Immunoglobulin levels were compared in 287 individually sex and age matched SLE patients and 287 population controls (A, C, E, G, I). IgM and IgG levels were also compared in different SLE patient subgroups based on their IgG autoantibody profiles (B, D, F, H, J, see Table 1 for subgroup criteria). Briefly, 318 population controls were compared to 78 SLE patients with antiphospholipid (APS) autoantibody profile (with IgG anti-CL/$\beta_2$GPI), 98 patients with Sjögren's syndrome (SS) autoantibody profile (with IgG anti-Ro/La) and the remaining patient group consisting of 160 SLE patients with other autoantibodies (Other Abs). Specific natural IgM levels, IgM anti-PC-BSA, IgM anti-CWPS and IgM anti-malondialdehyde (MDA) modified protein, were measured by ELISA and quantitative reactivates were depicted as Relative Units (RU)/ml. P-values were derived from Mann-Whitney or Kruskal-Wallis test, adjusted for multiple comparisons.

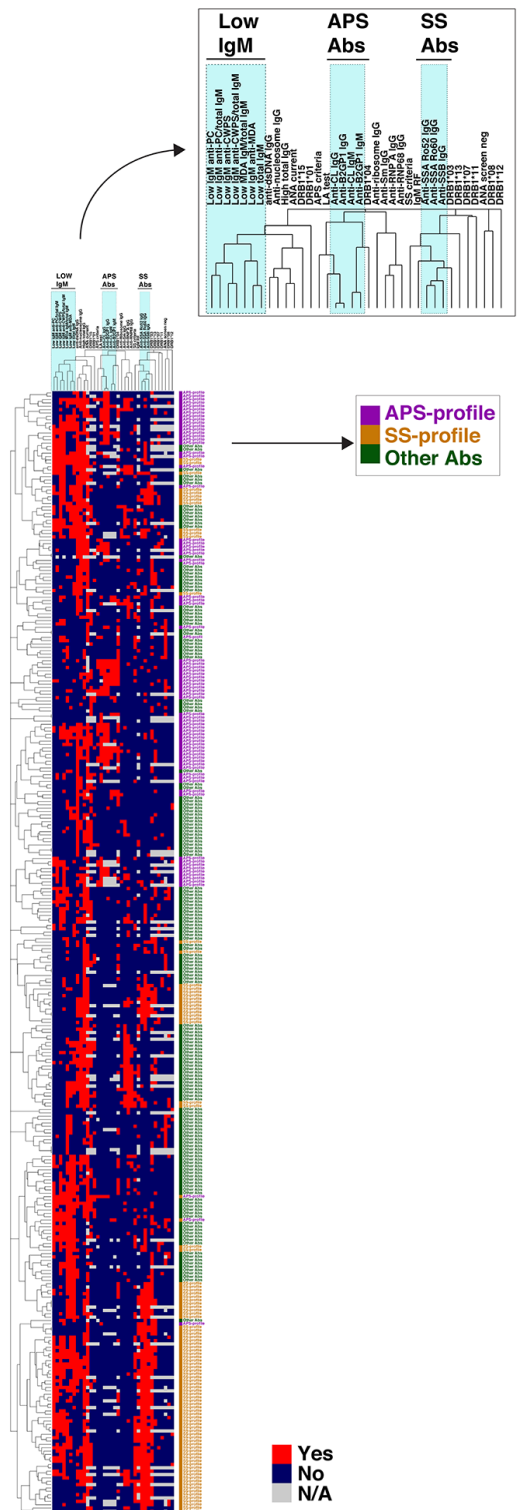

**Figure 2. Low IgM levels contribute to SLE heterogeneity**
Visualization of heterogeneity of SLE patients with regards to autoantibody reactivity, HLA-DRB1 positivity, and low IgM levels, using hierarchical uncentered cluster analysis of 336 SLE patients. All included variables were dichotomous (yes or no) and the data vas clustered both based on patient (y-axis) and variable (x-axis). The analysis shows for example that APS autoantibody positivity cluster with HLA-DRB1*04 and SS autoantibody positivity cluster with HLA-DRB1*03. The SLE patients that were categorized as having an SS-profile (yellow) or APS-profile (purple) primarily cluster together in separate clusters and the low IgM levels contribute to the patient heterogeneity and subgroup clustering. Cluster analysis was performed using Cluster 3.0 and visualization using Java Treeview.

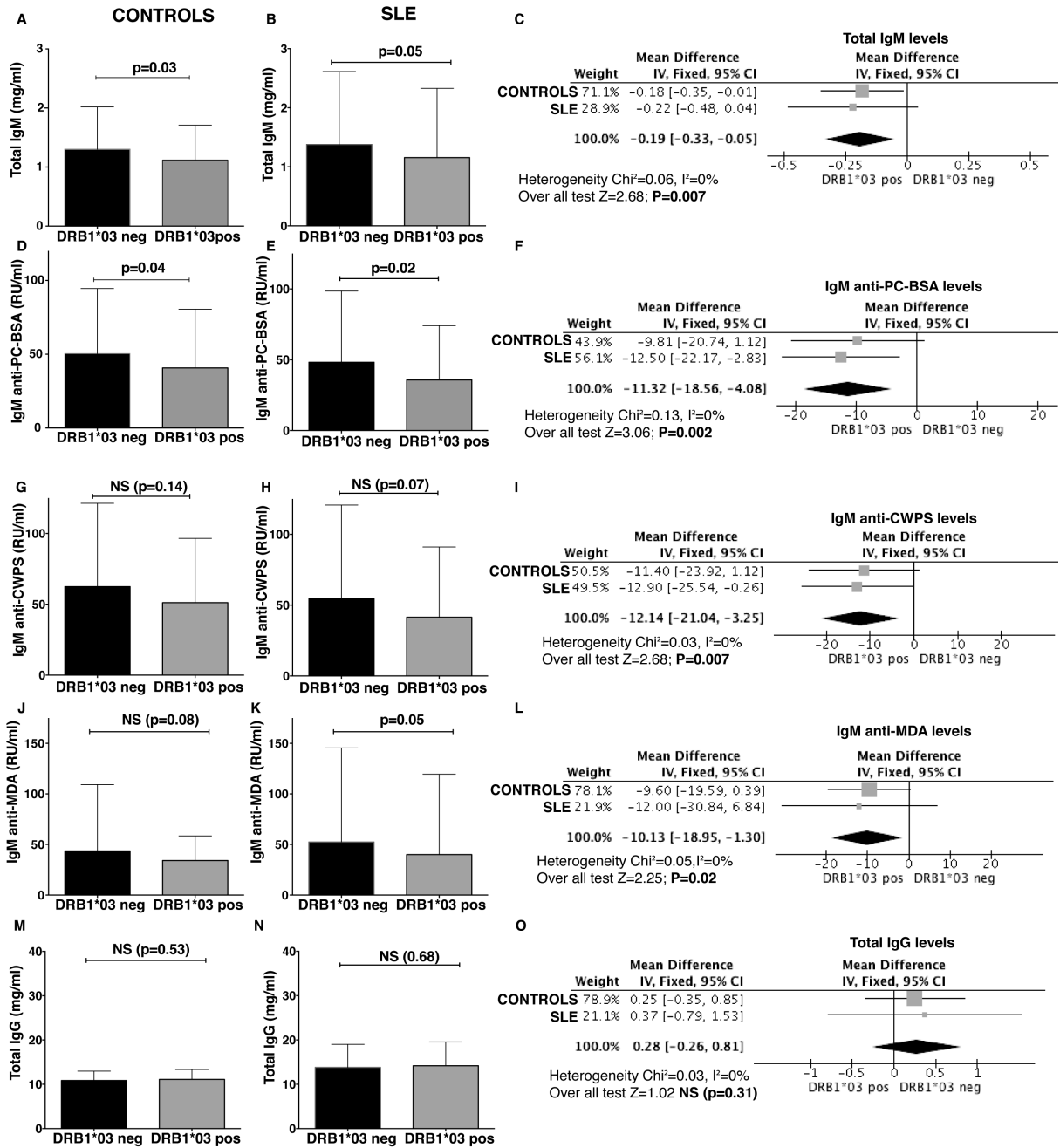

**Figure 3. Individuals with HLA-DRB1*03 genotype have lower levels of circulating IgM independent of autoreactivity**

The left panels show immunoglobulin levels in 249 HLA-DRB1*03 negative population controls compared to 68 HLA-DRB1*03 positive individuals, and 174 HLA-DRB1*03 negative SLE patients compared to 150 HLA-DRB1*03 positive SLE patients. The DRB1*03 positive groups included individuals that were both heterozygous and homozygous for the allele. The presented p-values were calculated using Mann-Whitney analysis. The right panels show the meta-analysis combining controls and the SLE groups, and calculating the difference in means using Inverse Variance method with fixed effect. Forest plots visualizing the mean differences (DRB1*03 positive – DRB1*03 negative) in the different cohorts and the combined data are presented (adapted from Review Manager 5.3 outputs).

**Table 1. Division of SLE patients into subgroups based on IgG autoantibody profiles**

| SLE with SS-profile (N=78) | SLE with APS-profile (N=98) | SLE with other autoAbs# (N=160) |
|---|---|---|
| IgG anti-Ro52/Ro60/La $\geq$2 positive tests | IgG anti-CL/$\beta_2$GPI = 2 positive tests | Remaining SLE |
| LA negative | *and/or* Lupus Anticoagulant positive test | IgG Anti-Ro52/Ro60/La $\leq$1 positive test |
| | *and/or* IgM/IgA anti-CL/$\beta_2$GPI $\geq$2 positive tests | LA negative |

# This group includes patients with anti-RNP/Sm autoantibodies but also subjects who were both ANA-negative/APS-negative, and the ANA-positive SLE patients with only IgG anti-dsDNA.

**Table 2. IgM levels in SLE patients compared to matched population controls**

|  | Controls * (N=287) | SLE (N=287) |  |
|---|---|---|---|
|  | Frequency/ Mean | Frequency/ Mean | p-value** |
| Age (years; Mean±SD) | 47.9±14.7 | 47.7±14.7 | NS (0.87) |
| Female (% n/N)[#] | 92% (265/287) | 92% (265/287) | NS (1) |
| **Antibody levels** |  |  |  |
| Total IgM (mg/ml; Mean±SD [median]) | 1.265±0.70 [1.00] | 1.276±1.1 [0.96] | **0.02** |
| Low total IgM (<0.8mg/ml)[§] | 25% (71/283) | 38% (109/283) | **0.0008** |
| α-PC-BSA IgM (RU/ml; Mean±SD [median]) | 48.1±43.1 [33.3] | 43.4±46.6 [24.4] | **0.05** |
| Low α-PC-BSA IgM (<14 RU/ml)[§] | 26% (75/287) | 29% (84/287) | NS (0.3) |
| α-CWPS IgM (RU/ml; Mean±SD [median]) | 59.7±56.4 [36.2] | 49.5±61.4 [26.8] | **<0.0001** |
| Low α-CWPS IgM (<21 RU/ml)[§] | 25% (73/287) | 40% (115/287) | **0.0003** |
| α-MDA IgM (RU/ml; Mean±SD [median]) | 41.8±62 [31.0] | 46.2±76 [25.4] | **0.009** |
| Low α-MDA IgM (<22 RU/ml) [§] | 25% (74/287) | 41% (118/287) | **0.0001** |
| Total IgG (mg/ml; Mean±SD [median]) | 10.9±2.0 [10.9] | 14.1±5.2 [14.3] | **<0.0001** |
| High total IgG (>12mg/ml)[§§] | 25% (70/283) | 60% (170/283) | **<0.0001** |

n= number of patients with a positive test
N= number of patients with measurement information
SD= standard deviation
*Population controls, age and sex matched to SLE patients.
n= 287 for anti-PC and anti-MDA levels; n=283 for total IgM and total IgG levels
** p-values derived from Mann-Whitney test or Fisher's exact test
§ Cutoff based on lowest quartile (25%) of all available controls N=322 (anti-PC/MDA) N=318 (total IgM)
§§ Cutoff based on highest quartile (75%) of all available controls N=318
Patients ever receiving Rituximab treatment were excluded

**Table 3. Patient characteristics and IgM levels in different SLE subgroups**

| | Controls (N=318) | SLE with APS-profile (N=78) | | SLE with SS-profile (N=98) | | SLE with other autoAbs (N=160) | | *APS vs SS profile* |
|---|---|---|---|---|---|---|---|---|
| | Frequency/ Mean | Frequency/ Mean | p-value* | Frequency/ Mean | p-value* | Frequency/ Mean | p-value* | p-value** |
| **Patient characteristics** | | | | | | | | |
| Age (years, Mean ±SD) | 47.4±15 | 49.5±15 | NS (0.89) | 48.7±14.4 | NS (0.99) | 45.4±16 | NS (0.45) | NS (0.73) |
| Female | 92% (292/318) | 89% (69/78) | NS (0.36) | 85% (83/98) | 0.05 | 94% (150/160) | NS (0.58) | NS (0.51) |
| SLEDAI ≥6 (%, n/N) | N/A | 39% (29/75) | N/A | 29% (28/97) | N/A | 27% (41/150) | N/A | NS (0.19) |
| APS syndrome# % | N/A | 51% (39/77) | N/A | 3% (3/96) | N/A | 4% (7/160) | N/A | <0.0001 |
| Sjögren's syndrome# % | N/A | 14% (11/78) | N/A | 41% (40/97) | N/A | 19% (31/160) | N/A | <0.0001 |
| HLA DRB1 *03 pos (%, n/N) | 21.4% (67/313) | 26% (16/62) | NS (0.50) | 79% (67/85) | <0.0001 | 37% (49/134) | 0.001 | <0.0001 |
| HLA DRB1 *04 pos (%, n/N) | 33.2% (107/313) | 57% (35/62) | 0.002 | 18% (15/85) | 0.003 | 27% (36/134) | NS (0.15) | <0.0001 |
| HLA DRB1 *15 pos (%, n/N) | 29.7% (93/313) | 31% (19/62) | NS (0.88) | 27% (23/85) | NS (0.68) | 47% (63/134) | 0.0005 | NS (0.71) |
| **SLE autoantibodies** | | | | | | | | |
| Total IgG (mg/ml; Mean±SD) | 10.9±2.1 | 14±5.1 | <0.0001 | 16±6.6 | <0.0001 | 13±4.7 | <0.0001 | 0.005 |
| High total IgG (>12mg/ml) | 27% (87/318) | 55% (43/78) | <0.0001 | 74% (72/98) | <0.0001 | 54% (85/159) | <0.0001 | 0.02 |
| $\alpha$-dsDNA IgG pos (%, n/N)∈ | 1.4% (4/278) | 50% (39/78) | <0.0001 | 28% (27/98) | <0.0001 | 38% (61/160) | <0.0001 | 0.005 |
| $\alpha$-CL IgG pos (%, n/N) | 0% (0/275) | 76% (50/66) | <0.0001 | 0% (0/82) | NS (1) | 0% (0/158) | NS (1) | <0.0001 |
| $\alpha$-$\beta_2$GPI IgG pos (%, n/N) | 0% (0/273) | 74% (49/66) | <0.0001 | 0% (0/82) | NS (1) | 3% (4/158) | 0.02 | <0.0001 |
| $\alpha$-Ro52/SSA IgG pos (%, n/N) | 1.1% (3/274) | 5% (4/78) | 0.05 | 90% (88/98) | <0.0001 | 3% (4/160) | NS (0.43) | <0.0001 |
| $\alpha$-Ro60/SSA IgG pos (%, n/N) | 1.8% (5/274) | 13% (10/78) | 0.0002 | 100% (98/98) | <0.0001 | 18% (28/160) | <0.0001 | <0.0001 |
| $\alpha$-La/SSB IgG pos (%, n/N) | 2.9% (8/274) | 3% (2/78) | NS (1) | 79% (77/98) | <0.0001 | 2% (3/160) | NS (0.75) | <0.0001 |
| $\alpha$-Sm IgG pos (%, n/N) | 0.4% (1/274) | 14% (11/78) | <0.0001 | 12% (12/98) | <0.0001 | 25% (40/160) | <0.0001 | NS (0.82) |
| $\alpha$-RNPa IgG pos (%, n/N) | 3.2% (9/274) | 17% (13/78) | 0.0001 | 15% (15/98) | 0.0001 | 31% (50/160) | <0.0001 | NS (0.84) |
| RF IgM pos (%, n/N) | 4.9% (14/283) | 10% (6/58) | NS (0.12) | 51% (41/81) | <0.0001 | 19% (25/132) | <0.0001 | <0.0001 |
| RF IgG pos (%, n/N) | 3.8% (10/261) | 18% (10/57) | 0.0007 | 29% (21/72) | <0.0001 | 10% (12/120) | 0.03 | NS (0.15) |
| **IgM natural antibodies** | | | | | | | | |
| Total IgM (mg/ml; Mean±SD) | 1.26±0.68 | 1.6±1.7 | NS (0.99) | 0.97±0.58 | <0.0001 | 1.27±1.1 | NS (0.09) | 0.0004 |
| Low total IgM (<0.8 mg/ml)§ | 25% (79/318) | 28% (22/78) | NS (0.56) | 45% (44/98) | 0.0002 | 38% (60/159) | 0.004 | 0.03 |
| $\alpha$-PC-BSA IgM (RU/ml; Mean±SD) | 48.5±43 | 54±54 | NS (0.99) | 28±32 | <0.0001 | 46±45 | NS (0.99) | 0.0005 |
| Low $\alpha$-PC-BSA IgM (<14 RU/ml)§ | 25% (79/318) | 24% (19/78) | NS (1) | 46% (45/98) | 0.0001 | 23% (37/160) | NS (0.74) | 0.004 |
| $\alpha$-CWPS IgM (RU/ml; Mean±SD) | 60.7±57 | 61±80 | NS (0.17) | 32±35 | <0.0001 | 50±54 | 0.009 | 0.02 |
| Low $\alpha$-CWPS IgM (<21 RU/ml)§ | 24% (77/318) | 34% (27/78) | NS (0.08) | 47% (46/98) | <0.0001 | 36% (57/160) | 0.01 | NS (0.12) |
| $\alpha$-MDA IgM (RU/ml; Mean±SD) | 42±59 | 87±157 | NS (0.08) | 32±30 | 0.003 | 39±44 | 0.04 | 0.0001 |
| Low a-MDA IgM (<22 RU/ml)§ | 25% (81/322) | 23% (18/78) | NS (0.77) | 47% (46/98) | <0.0001 | 39% (63/160) | 0.002 | 0.002 |

n= number of patients with a positive test

N= number of patients with measurement information

SD= standard deviation

# Fulfill clinical criteria for secondary Antiphospholipid syndrome (APS) by Miyakis [6] or Sjögren's syndrome by the European/American consensus [9]

* P-value compared to controls from Kruskal-Wallis test, adjusted for multiple comparisons, or Fisher's exact test

§ Cutoff based on lowest quartile (25[th] percentile) of all available controls N=322 (anti-PC/MDA) n=318 (total IgG/IgM)

§§ Cutoff based on highest quartile (75[th] percentile) of all available controls N=318

€ In the complete SLE cohort we observed 38% IgG anti-dsDNA positivity

** P-values derived from Mann-Whitney test or Fisher's exact test

Patients that had received Rituximab treatment were excluded

**Table 4. IgM levels are lower in individuals positive for HLA-DRB1*03**

§ Cutoff based on lowest quartile (25[th] percentile) of all available controls N=322

| | Controls | | | SLE | | | Meta-analysis: Controls and SLE | |
|---|---|---|---|---|---|---|---|---|
| | DRB1*03 neg (n=249) | DRB1*03 pos (n=68) | | DRB1*03 neg (n=174) | DRB1*03 pos (n=150) | | DRB1*03 pos vs DRB1*03 neg | |
| | Frequency/ Mean±SD | Frequency/ Mean±SD | P-value# | Frequency/ Mean±SD | Frequency/ Mean±SD | P-value# | Mean difference/ Odds Ratio, [95% CI] | P-value ¶ |
| Total IgM (mg/ml) | 1.30±0.72 | 1.12±0.59 | **0.03** | 1.38±1.24 | 1.16±1.17 | **0.05** | MD: -0.19 [-0.33, -0.05] | **0.007** |
| Low Total IgM§ | 23% (57/246) | 33% (22/67) | NS (0.11) | 34% (60/174) | 42% (63/150) | NS (0.17) | OR: 1.46 [1.02, 2.09] | **0.04** |
| α-PC-BSA IgM (RU/ml) | 50.5±44.0 | 40.7±39.8 | **0.04** | 48.2±50.3 | 35.7±38.3 | **0.02** | MD: -11.32 [-18.56, -4.08] | **0.002** |
| Low α-PC-BSA IgM§ | 21% (53/249) | 40% (27/68) | **0.003** | 26% (45/174) | 37% (55/150) | **0.04** | OR: 1.92 [1.33, 2.77] | **0.0004** |
| α-CWPS IgM (RU/ml) | 62.5±58.8 | 51.1±45.4 | NS (0.14) | 54.6±66.1 | 41.7±49.7 | NS (0.07) | MD: -12.14 [-21.04, -3.25] | **0.007** |
| Low α-CWPS IgM§ | 22% (56/249) | 32% (22/68) | NS (0.11) | 36% (63/174) | 44%( 66/150) | NS (0.17) | OR: 1.47 [1.03, 2.10] | **0.03** |
| α-MDA IgM (RU/ml) | 43.7±66 | 34.1±24 | NS (p=0.08) | 52.0±93 | 40.0±80 | **0.05** | MD: -10.13 [-18.95, -1.30] | **0.02** |
| Low α-MDA IgM § | 25% (62/249) | 26% (18/68) | NS (p=0.87) | 37% (64/174) | 48% (73/150) | **0.03** | OR: 1.42 [0.99, 2.02] | NS (0.06) |
| Total IgG (mg/ml) | 10.8±2.1 | 11.1±2.2 | NS (0.53) | 13.8±5.2 | 14.2±5.4 | NS (0.68) | MD: 0.28 [-0.26, 0.81] | NS (0.31) |
| High total IgG§§ | 24% (58/246) | 30% (20/67) | NS (0.34) | 58% (101/174) | 58% (87/150) | NS (1) | OR: 1.12 [0.78, 1.59] | NS (0.55) |

§§ Cutoff based on highest quartile (75[th] percentile) of all available controls N=318
# p-value from 2-sided Mann-Whitney test or Fisher's exact test
¶ p-value from Meta-analysis using Mantel-Haenszel method (for frequencies) for Inverse Variance method (for comparing means) with fixed effect.
Patients ever receiving Rituximab treatment were excluded from the analysis.
NS= Non-significant, CI= confidence interval, MD=Mean difference, OR= Odds ratio
HLA-DRB1*03 positive patients, includes patients both heterozygous and homozygous for the allele.

**Table 5. Cardiovascular risk in association to low IgM anti-PC in different SLE subgroups**

| | All SLE | | | High IgM anti-PC-BSA >14 RU/ml§ | Low IgM anti-PC-BSA ≤14 RU/ml | | | High IgM anti-CWPS >16 RU/ml§ | Low IgM anti-CWPS ≤16 RU/ml | | |
|---|---|---|---|---|---|---|---|---|---|---|---|
| | Frequency | OR [95% CI] | p-value** | Frequency | Frequency | OR [95% CI] | p-value# | Frequency | Frequency | OR [95% CI] | p-value# |
| **All SLE** | | | | | | | | | | | |
| Plaque€ | 21% (55/263) | | | 16% (31/188) | 31% (24/80) | **2.2 [1.2-4.0]** | **0.02** | 12% (20/162) | 33% (35/106) | **3.5 [1.9-6.5]** | **<0.0001** |
| Vascular events* | 26% (98/380) | | | 23% (61/263) | 32% (37/117) | 1.5 [0.94-2.5] | NS (0.10) | 23% (53/230) | 30% (45/150) | 1.4 [0-90-2.3] | NS (0.15) |
| **SLE with APS-profile** | | | | | | | | | | | |
| Plaque | 25% (13/52) | 1.3 [0.62-2.5] | NS (0.58) | 20% (8/39) | 39% (5/13) | 2.4 [0.62-9.5] | NS (0.26) | 24% 8/34 | 27% (5/18) | 1.25 [0.34-4.6] | NS (0.75) |
| Vascular events | 48% (36/75) | **2.7 [1.6-4.4]** | **0.0003** | 47% (27/57) | 50% (9/18) | 1.1 [0.38-3.2] | NS (1.0) | 47% (23/49) | 50%(13/26) | 1.1 [0.4-2.9] | 0.81 |
| **SLE with SS-profile** | | | | | | | | | | | |
| Plaque | 16% (12/73) | 0.74 [0.37-1.5] | NS (0.51) | 5% 2/38 | 28% (10/35) | **7.2 [1.5-36]** | **0.01** | 3% (1/36) | 30% (11/37) | **14.8 [1.8-122]** | **0.003** |
| Vascular events | 14% (13/94) | **0.46 [0.24-0.87]** | **0.01** | 14% (7/51) | 14% (6/43) | 1.0 [0.31-3.3] | NS (1.0) | 12% (6/50) | 16% (7/44) | 1.39 [0.4-4.5] | NS (0.77) |
| **SLE with other autoAbs** | | | | | | | | | | | |
| Plaque | 23% (28/123) | 1.1 [0.67-1.9] | NS (0.69) | 19% (19/98) | 36% (9/25) | 2.3 [0.90-6.1] | NS (0.11) | 14% (11/80) | 40% (17/43) | **4.1 [1.7-9.9]** | **0.002** |
| Vascular events | 22% (34/153) | 0.82 [0.52-1.3] | NS (0.44) | 18% (21/116) | 35% (13/37) | **2.5 [1.1-5.6]** | **0.04** | 19% (18/97) | 29% (16/56) | 1.8 [0.81-3.8] | NS (0.16) |

*History of vascular events, includes both arterial and venous events
** p-value by Fisher's exact test for the subgroup compared to all SLE
§ Cutoff based on lowest quartile (25[th] percentile) of all available controls n=322
€ Presence of atherosclerotic plaque by carotid ultrasound intima media thickness measurements
#p-value by Fisher's exact test for low IgM levels compared to high IgM levels (above cutoff for low levels).
OR: Odds ratio; CI: Confidence interval